\documentclass[runningheads]{llncs}
\usepackage{graphicx}
\usepackage{amssymb}
\usepackage{amsmath}
\usepackage{hyperref}
\usepackage{subfiles}
\def\*#1{\mathbf{#1}}

\begin{document}
\title{Uncertainty Estimation and Propagation in Accelerated MRI Reconstruction}%\thanks{Supported by organization x.}}
%\titlerunning{Uncertainty Estimation and Propagation in Accelerated MRI}
%
%\titlerunning{Abbreviated paper title}
% If the paper title is too long for the running head, you can set
% an abbreviated paper title here
%
\author{Paul Fischer\inst{1} \and Thomas Küstner\inst{2} \and
Christian F. Baumgartner\inst{1}}
\authorrunning{Fischer et al.}
% First names are abbreviated in the running head.
% If there are more than two authors, 'et al.' is used.
%
\institute{Cluster of Excellence -- Machine Learning for Science, University of Tübingen \and Medical Image and Data Analysis Lab, University
Hospital of Tübingen}
%\email{lncs@springer.com}}
%
\maketitle              % typeset the header of the contribution
\begin{abstract}

MRI reconstruction techniques based on deep learning have led to unprecedented reconstruction quality especially in highly accelerated settings. However, deep learning techniques are also known to fail unexpectedly and hallucinate structures. This is particularly problematic if reconstructions are directly used for downstream tasks such as real-time treatment guidance or automated extraction of clinical paramters (e.g. via segmentation). Well-calibrated uncertainty quantification will be a key ingredient for safe use of this technology in clinical practice. In this paper we propose a novel probabilistic reconstruction technique (PHiRec) building on the idea of conditional hierarchical variational autoencoders. We demonstrate that our proposed method produces high-quality reconstructions as well as uncertainty quantification that is substantially better calibrated than several strong baselines. We furthermore demonstrate how uncertainties arising in the MR reconstruction can be propagated to a downstream segmentation task, and show that PHiRec also allows well-calibrated estimation of segmentation uncertainties that originated in the MR reconstruction process. 
\end{abstract}

\section{Introduction}
Fast magnetic resonance imaging (MRI) techniques play a vital role in clinical practice, allowing to scan an increased number of patients while alleviating patient discomfort caused by prolonged acquisition times. Highly accelerated MR acquisitions also hold the key unlocking novel applications such as real-time MR-guided radiation therapy~\cite{waddington_real-time_2022}, or shortened scans for directly estimating clinical parameters via segmentations potentially without human oversight~\cite{tolpadi2023k2s,caliva2020breaking,schlemper2018cardiac}. 

In recent years, MRI reconstruction techniques relying on deep learning (DL) have gained substantial interest due to their excellent performance at very high acceleration rates~\cite{ongie_deep_2020,jin_deep_2017,jalal_robust_2021} and ability to provide real-time reconstructions~\cite{zhou2019parallel,hauptmann2019real}. Although DL-based reconstructions often appear realistic and of high quality, they have also been shown to fail unexpectedly~\cite{gottschling2020troublesome}, and hallucinate structural details~\cite{morshuis2022adversarial}. Crucially, they lack the ability to indicate regions in the reconstructed images that are uncertain. This problem is exacerbated in scenarios where reconstructed images are used directly for downstream tasks, such as real-time treatment guidance~\cite{waddington_real-time_2022}, or extraction of clinical parameters via segmentation~\cite{tolpadi2023k2s}. 

The reconstruction process inherently contains \emph{aleotoric}, or irreducible, uncertainty due to the fact that a single undersampled acquisition can correspond to an infinite number of potential reconstructions with varying likelihoods~ \cite{zeng_review_2021}. Moreover, \emph{epistemic} uncertainty may arise when the reconstruction model is applied outside the domain on which it was trained. Developing DL-based reconstruction techniques that are able to reflect those uncertainties is crucial, especially in applications without human oversight.  

\begin{figure}[t]
    \centering
    \includegraphics[width=\textwidth]{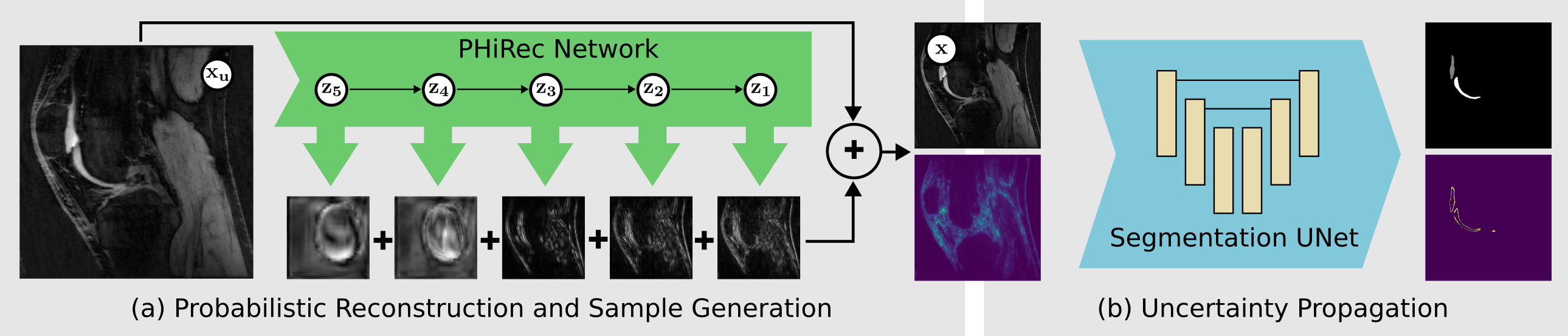}
    \caption{(a) In our proposed Probabilistic Hierarchical Reconstruction (PHiRec) model, five latent variables $\*z_l$ generate residual changes at different resolution scales. These residual changes are added to the undersampled input image $\*x_u$ to generate the final output $\*x$. The model can be used to sample likely reconstructions and to obtain reliable uncertainty estimates. (b) These samples can then be propagated to a subsequent segmentation network allowing to estimate the resulting segmentation uncertainty.}
    \label{fig:overview}
\end{figure}

% Related work
Several approaches have been proposed to model uncertainty of DL-based MRI reconstruction techniques. Schlemper et al.~\cite{schlemper2018bayesian} proposed to estimate epistemic and aleotoric uncertainty using MC dropout and a heteroscedastic variance term, respectively. Hepp et al.~\cite{hepp2022uncertainty} presented promising preliminary results for estimating epistemic uncertainty using an ensemble of networks. However, the approach does not scale well since the number of samples that can be generated is equal to the number of networks in the ensemble. Narnhofer et al.~\cite{narnhofer2021bayesian} modelled epistemic uncertainty by combining the previously proposed Total Deep Variation approach~\cite{kobler2020total} with the Bayes-by-Backprop technique, which models every network weight as a Gaussian with individual mean and variance~\cite{blundell2015weight}. We note that this approach has heavy GPU-memory requirements and limits the complexity of the network architecture that can be used. Tezcan et al.~\cite{tezcan2022sampling} have proposed learning the prior distribution of MR images using a variational autoencoder (VAE) and using Markov-Chain Monte-Carlo (MCMC) to sample possible reconstructions. However, this approach is severely limited by the sampling times and is not suitable for real-time applications. Angelopoulos et al. \cite{angelopoulos_image--image_2022} proposed an uncertainty quantification method based on conformal prediction, which offers a straightforward implementation and comes with mathematical guarantees. However, this method does not allow generating samples which could be used to explore potential reconstructions or propagate uncertainty to subsequent tasks. Recently diffusion models have demonstrated exceptional reconstruction performance \cite{jalal_robust_2021,peng-towards,chung2022scorebased,xie2022measurementconditioned}. While uncertainty quantification is feasible in those models, it is currently hindered by extremely long sampling times rendering them unsuitable for real-time applications. In closely related work on a different imaging modality, Zhang et al.~\cite{zhang2021conditional} proposed a PET reconstruction method based on conditional VAEs (cVAE)~\cite{sohn2015learning}. The approach employed a bottleneck architecture which only allows to model the uncertainty at a low spatial resolution and is prone to producing blurry reconstructions. Lastly, a major limitation of the existing literature is that none of the above studies present a quantitative evaluation of uncertainty quantification or a thorough comparison with baseline methods, instead relying solely on qualitative interpretation of the uncertainty maps.

\setcounter{footnote}{0} 
In this paper, we propose a novel approach for estimating aleotoric uncertainty based on a hierarchical conditional VAE~\cite{sohn2015learning}. Hierarchical cVAEs have been shown to perform exceptionally well for estimating aleotoric uncertainty in segmentation tasks~\cite{phiseg,kohl2019hierarchical}. Specifically, they address two issues encountered in non-hierarchical cVAEs: blurry samples and limited expressivity to model high-dimensional spatial probability distributions~\cite{punet,zhang2021conditional}. However, despite their promise, hierarchical cVAEs remain unexplored in the context of MRI reconstruction. Here, we build on the Probabilistic Hierarchical Segmentation (PHiSeg) model by Baumgartner et al.~\cite{phiseg}, which was originally proposed for segmentation to create a novel \emph{P}robabilistic {\emph{Hi}erarchical \emph{R}econstruction technique which we coin PHiRec\footnote{The code for PHiRec is available at \url{https://github.com/paulkogni/MR-Recon-UQ}}. Our contributions are as follows:
\begin{itemize}
    \item We propose PHiRec and show that it outperforms several strong baselines in terms of calibration of its uncertainty quantification.
    \item We demonstrate that the uncertainties originating in the MR reconstruction can be \emph{propagated} to a downstream segmentation task in order to estimate the resulting segmentation uncertainties. This is, to the best of our knowledge, the first work to explore the propagation of uncertainties arising in DL-based MRI reconstruction to a downstream task. 
    \item We present the first comprehensive \emph{quantitative} evaluation of uncertainty quantification for MRI reconstruction contrasting several baselines. 
\end{itemize}

\section{Methods}
We denote a fully-sampled MR image as $\*x \in \mathbb{C}^N$ where $N$ is the number of pixels. In a multi-coil MR acquisition, the acquired $k$-space data can be modelled as $\*y = \mathcal{M}\mathcal{F}\mathcal{S}\*x + \*\eta$, where $\mathcal{M}$ is an undersampling operator, $\mathcal{F}$ is the Fourier operator, $\mathcal{S}$ is an operator encoding the spatial sensitivity of each coil, and $\*\eta$ is used to model thermal scanner noise. The goal of MRI reconstruction is to estimate the maximimum a-posteriori of the distribution $p(\*x|\*y)$. For uncertainty estimation we are additionally interested in the spread of this distribution.  

We pose the reconstruction as a de-aliasing problem by modelling the distribution $p(\*x|\*x_u)$, where the undersampled image $\*x_u$ is obtained by applying the inverse Fourier operator to the zero-filled measurement data $\*y$. In the following, we show how a hierarchical cVAE approach can be employed to model the distribution $p(\*x|\*x_u)$. As shown in Fig.\,\ref{fig:overview}, we model the distribution using a cVAE that has $L=5$ separate latent variables $\*z_l$ each operating on a different resolution scale. For instance, $\*z_1$ operates at the original image resolution, while $\*z_5$ operates at a resolution that was four times downpooled by a factor of 2. Each resolution level is responsible for probabilistically generating residual changes that are added to the input image $\*x_u$ in order to remove undersampling artifacts and obtain the reconstructed image $\*x$. Our modelling assumption includes that the distribution of each $\*z_l$ depends on the input image $\*x_u$ as well as the latent variable of the resolution level below $\*z_{l+1}$. This allows higher resolution levels to have a notion of what changes were already performed in the resolution level below. As was previously shown, this hierarchical approach is a very expressive model for capturing high-dimensional probability distributions~\cite{phiseg,kohl2019hierarchical}. Note that in contrast to the hierarchical cVAE methods developed in the context of segmentation~\cite{kohl2019hierarchical,phiseg}, our proposed PHiRec model contains a skip connection from the input to the output which we found to facilitate the de-aliasing problem. This is reflected by the dependence of the likelihood  $p(\*x | \*z_{1:L}, \*x_u)$ on $\*x_u$ in the equations below.

Using the above modelling assumptions $p(\*x | \*x_u)$ can be written as
\begin{equation*}
\label{eq:distr}
    p(\*x | \*x_u) = \int p(\*x | \*z_{1:L}, \*x_u) p(\*z_1 | \*z_2, \*x_u) \dots p(\*z_{L-1} | \*z_L, \*x_u) p(\*z_L | \*x_u) d\*z_{1:L}.
\end{equation*}
Following the standard variational approach we maximise the evidence lower bound, $ \text{ELBO}(\*x|\*x_u) := \log p(\*x|\*x_u) - KL(q(\*z_{1:L}|\*x, \*x_u)||(p(\*z_{1:L}|\*x, \*x_u))$, which is a lower bound on the true log likelihood.  Using our model assumptions, and following the derivation in Baumgartner et al.~\cite{phiseg}, we can write the ELBO as
\begin{equation*} \label{eq:elbo}
\begin{split}
\text{ELBO}(\*x|\*x_u) = & \mathbb{E}_{q(\*z_{1:L} | \*x_u, \*x) } \left[ \log p(\*x | \*z_{1:L}, \*x_u)  \right] - \alpha_L KL\left[ q(\*z_L | \*x, \*x_u) || p(\*z_L | \*x_u) \right] \\
    & - \sum_{l=1}^L \alpha_l \mathbb{E}_{q(\*z_{l+1} | \*x_u, \*x)} \left[  KL\left[ q(\*z_l | \*z_{l+1}, \*x, \*x_u) || p(\*z_l | \*z_{l+1}, \*x_u) \right]  \right],
\end{split}
\end{equation*}
where $\alpha_l := 4^{(l-1)}$ are heuristic weight terms to equalize the magnitude of the KL-terms of the different resolution levels. The prior and posterior distributions are modelled using axis-aligned Normal distributions
\begin{equation*}
\begin{split}
    p(\*z_l | \*z_{l+1}, \*x_u) = & \mathcal{N} \left(\*z_l | \Phi^{(\mu)}_l (\*z_{l+1}, \*x_u), \Phi^{(\sigma)}_l (\*z_{l+1}, \*x_u) \right) \\
    q(\*z_l | \*z_{l+1}, \*x, \*x_u) = & \mathcal{N} \left(\*z_l | \Theta^{(\mu)}_l (\*z_{l+1}, \*x, \*x_u), \Theta^{(\sigma)}_l (\*z_{l+1}, \*x, \*x_u) \right),
\end{split}
\end{equation*}
where $\Phi^{(\mu)}_l$, $\Phi^{(\sigma)}_l$, $\Theta^{(\mu)}_l, \Theta^{(\sigma)}_l$ are neural network functions that estimate each distribution's mean and variance. The likelihood of the final de-aliased reconstruction $p(\*x | \*z_{1:L}, \*x_u)$ is also modelled as a Normal distribution with a fixed variance and a mean that is estimated using another neural network. While in principle a mathematically valid model could be implemented using any neural network architecture the method lends itself to implementation as a U-Net-like architecture with the prior and posteriors implemented as U-Net encoders, and the likelihood as a decoder. For simplicity here we use the architecture proposed by Baumgartner et al.~\cite{phiseg} with the aforementioned addition of a skip connection from input to output (see Fig.~\ref{fig:overview}). A schematic of this architecture is shown in the supplementary materials. 

We train the entire architecture end-to-end with pairs of undersampled images $\*x_u$ and ground truth reconstructions $\*x$ using the ELBO defined above as objective. After training the posterior network is no longer required. The prior network can be used to predict the means and standard deviations of the $\*z_l$ variables. Given these values an arbitrary number of latent variable samples can be generated, and decoded using the likelihood network, to obtain final reconstruction samples. The mean prediction as well as the spread of the distribution $p(\*x|\*x_u)$ can then be calculated from these samples $\{\*x_i\}$. \\

\noindent \textbf{Uncertainty Propagation} Given a separately trained deterministic segmentation network $f: \*x \mapsto \*s$, we can furthermore estimate the distribution of the segmentations $\*s$ given the undersampled image $\*x_u$, $p(\*s|\*x_u)$, using the Monte Carlo method. Specifically, we can segment each of our reconstruction samples $\{\*x_i\}$ using $f$ and analyse the resulting distribution of segmentations empirically.

\section{Experiments and Results}

\noindent \textbf{Baselines} 
We compared our proposed PHiRec technique to several baseline strategies for estimating aleotoric and epistemic uncertainty. Firstly, we compared with Schlemper et al.'s~\cite{schlemper2018bayesian} approach for which we separately evaluated the epistemic uncertainty quantification based on MC Dropout,  the aleatoric uncertainty estimation rooted in a heteroscedastic variance term, as well as the combination of the two approaches as originally described. Furthermore, we compared to the ensemble based approach for estimating epistemic uncertainty initially demonstrated by Hepp et al.~\cite{hepp2022uncertainty}. Specifically, we created an ensemble of 20 separately trained reconstruction networks. Lastly, we extended the probabilistic U-Net~\cite{punet} to MRI reconstruction using the same strategy as for PHiRec to allow a comparison to another cVAE-based method. In order to focus the evaluation on the uncertainty quantification mechanism rather than on architectural details, all baseline methods were implemented using U-Nets, or U-Net-like architectures in the case of the probabilistic U-Net and our PHiRec. Furthermore, to ensure a fair comparison to our proposed approach, we implemented a skip connection from input to output for all baselines. 
%For all investigated baselines this either had no effect, or led to slight improvements. 
We used the mean and standard deviation calculated using 20 samples for all methods and in all experiments to obtain the final prediction and spread of the distribution, respectively. \\

\begin{figure}[t]
    \centering
    \includegraphics[width=\textwidth]{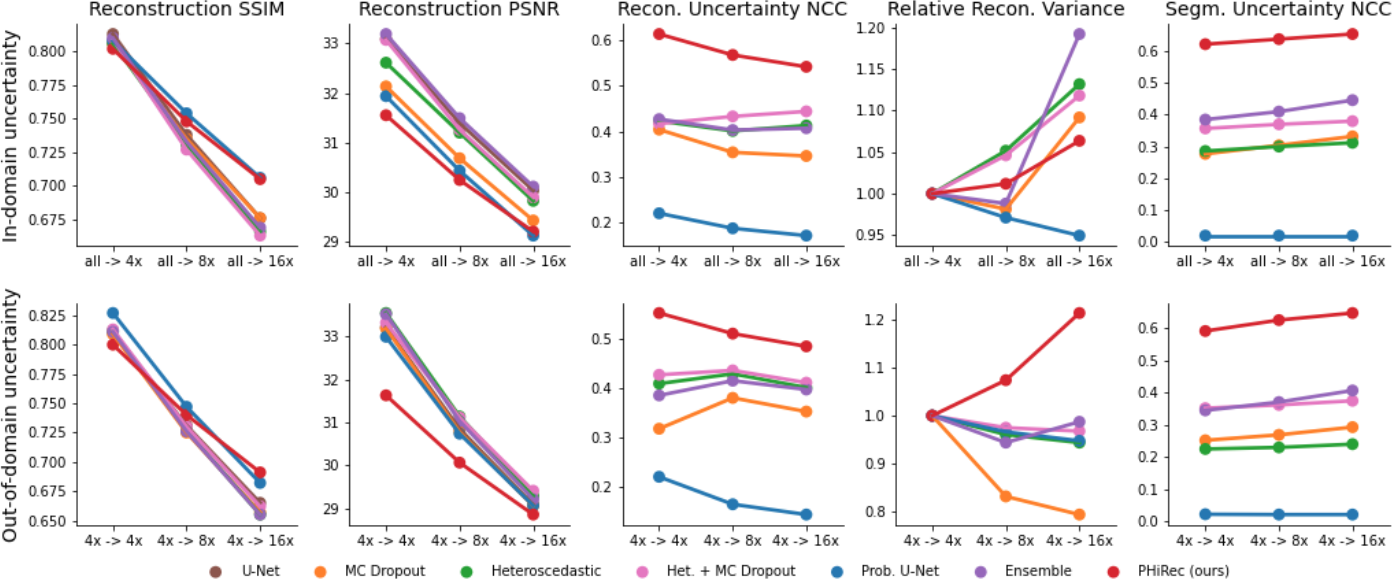}
    \caption{Quantitative results for the ID (top row), and OOD (bottom row) settings.}
    \label{fig:metrics}
\end{figure}

\noindent \textbf{Data} All experiments were performed on the Stanford Knee MRI Multi-Task Evaluation (SKM-TEA) dataset \cite{desai2022skmtea}, which comprises raw multi-coil $k$-space data of knee scans, along with segmentations for six anatomical structures. We employed the supplied undersampling masks, which were designed with a Poisson-Disc pattern. The provided coil sensitivities were used in a SENSE reconstruction \cite{pruessmann1999sense} to obtain the fully-sampled ground truth reconstruction $\*x$ as well as the undersampled network inputs $\*x_u$. We divided the dataset into a training, validation, and test set using the official splits.\\

\noindent \textbf{Experiment Settings} All experiments were performed in two distinct experimental settings: \emph{in-domain (ID)} and \emph{out-of-domain (OOD)}. In the ID setting, we simulatenously trained and also evaluated on images with acceleration rates 4x, 8x, and 16x. Since all acceleration factors have been seen during training, this setting is dominated by aleotoric uncertainty, that stems from the fact that there are multiple plausible solutions for each undersampled image. In the OOD setting, we trained only on images that have been accelerated 4x, but again tested on images with 4x, 8x, and 16x acceleration. In this setting, for 8x and 16x, there is an additional component of epistemic uncertainty in addition to the aleotoric uncertainty as the testing data moves away from the data that the model has seen during training.\\

\noindent \textbf{Training Details} All models were implemented in PyTorch \cite{NEURIPS2019_9015} and were trained with the Adam optimizer \cite{kingma2017adam} with a learning rate of $10^{-4}$ using a batch size of 6. The models were trained on NVIDIA RTX 2080 GPUs except the heteroschedastic models and PHiRec which were trained on an NVIDIA Tesla V100 GPU due to increased GPU memory demands. We trained all models for 10 days and model selection was performed based on structural similarity index (SSIM) of the reconstructions on a held-out validation set.\\

\noindent \textbf{Evaluation of Reconstruction Quality} For both the ID and OOD setting, we evaluated the reconstruction quality in terms of SSIM and peak signal to noise ratio (PSNR). Here, we additionally compared against a standard reconstruction U-Net without uncertainty quantification~\cite{hyun2018deep} to ensure the uncertainty quantification does not lead to a general performance degradation. The results are shown in the first two columns of Fig.\,\ref{fig:metrics}. We observed that, as expected, the performance of all methods degraded with increasing acceleration rates for both settings. This can also be visually confirmed by the squared error maps in Fig.\,\ref{fig:ood-images} for the OOD setting. Similar effects were observed for the ID setting (see results in supplementary materials). We further observed that all methods performed similarly in terms of pure reconstruction quality. However, PHiRec slightly underperformed in terms of PSNR, but slightly outperformed the other methods in terms of SSIM. This is consistent with the qualititative observation that, while PHiRec reproduced the structural properties exceptionally well, it had a slightly blurry quality. Example reconstructions for all methods are shown in the supplementary materials.\\

\begin{figure}[t]
    \centering
    \includegraphics[width=0.95\textwidth]{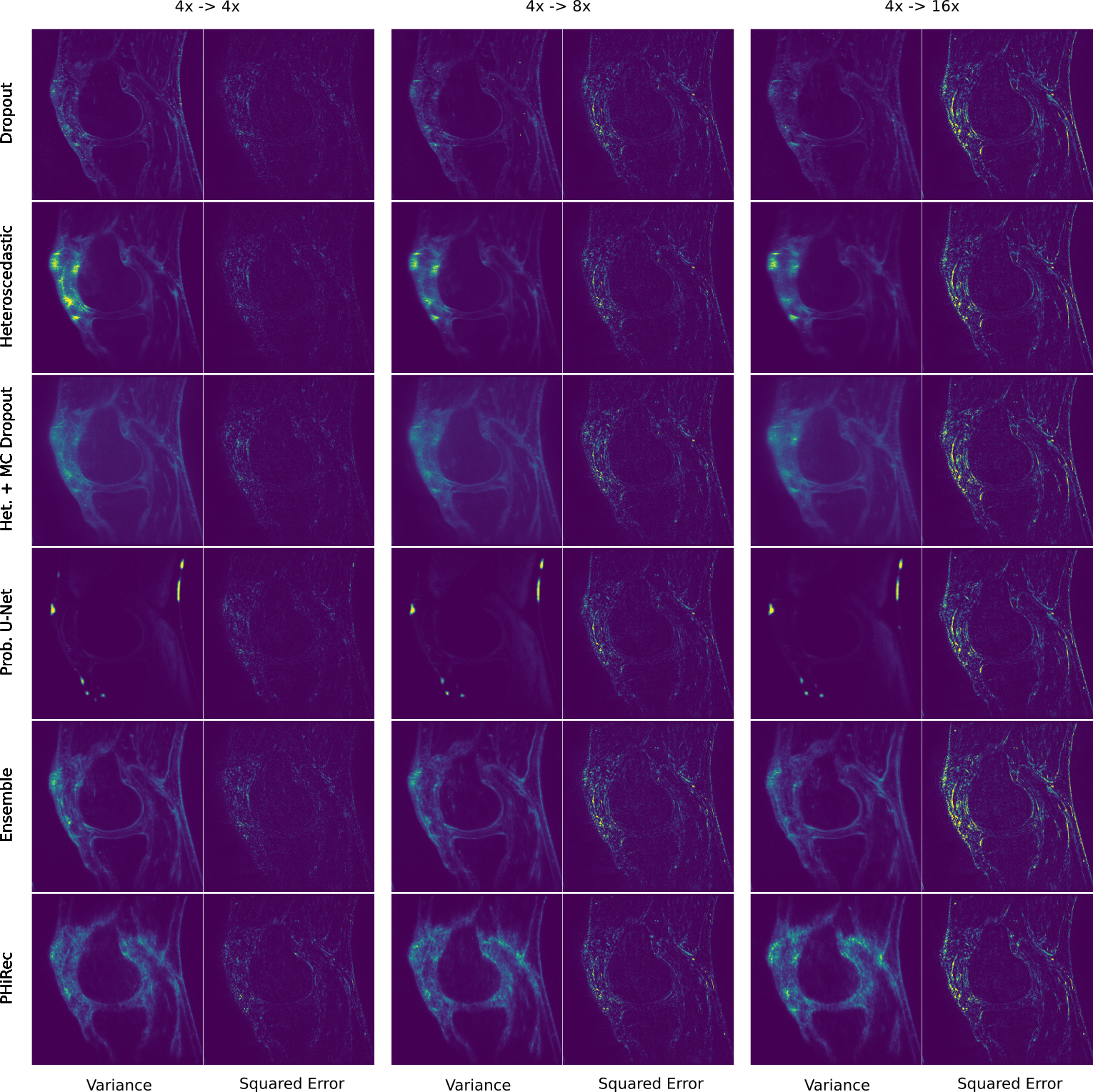}
    \caption{Variance maps and reconstruction squared error maps in the OOD setting where the column labels have the format ``train acceleration'' \textrightarrow \ ``test acceleration''.}
    \label{fig:ood-images} 
\end{figure}

\noindent \textbf{Evaluation of Reconstruction Uncertainty} A crucial quality for a robust uncertainty quantification method is that the model is \emph{calibrated}, i.e. that the uncertainty correlates with the model error~\cite{laves2021recalibration}. 
To assess calibration, we computed the average normalised cross correlation (NCC) between the reconstruction uncertainty and the reconstruction squared error for all test images. The results in the third column of Fig.\,\ref{fig:metrics} show that PHIRec is substantially better calibrated than the baselines in all settings. It is followed by the MC Dropout + Heteroscedastic Variance approach by Schlemper et al.~\cite{schlemper2018bayesian}. The probabilistic U-Net~\cite{punet}, which was originally proposed for segmentation in a multi-annotator regime, performed the worst in this category due to its poor sample diversity. These results can also be visually confirmed by comparing the uncertainty maps and corresponding error maps in Fig.\,\ref{fig:ood-images}. It is surprising that PHiRec, which is designed to model aleotoric uncertainties, also performed best in the OOD setting. We believe this might be due to the fact that differing acceleration factors do not constitute a large enough domain shift to add significant epistemic uncertainty. 

In addition to calibration, we also measured the intuition that the uncertainty should monotonically \emph{increase} with increasing acceleration rates. To this end, we calculated the relative change of the cumulative variance of all image pixels of the 8x and 16x settings with respect to the 4x setting. The results are shown in column four (``Relative Recon. Variance'') of Fig.\,\ref{fig:metrics}. Surprisingly, PHiRec is the only model for which the uncertainty consistently increases for higher acceleration rates. Instead, we found the uncertainty unexpectedly \emph{decreased} for most models between 4x and 8x acceleration. This can be visually confirmed for the OOD setting in the example image shown in Fig.\,\ref{fig:ood-images}. \\

\noindent \textbf{Evaluation Uncertainty Propagation} Lastly, we investigated propagating the uncertainty to a downstream segmentation task. To this end, we trained a standard deterministic segmentation U-Net with pairs of ground truth images $\*x$ and corresponding segmentation masks $\*s$ from the SKM-TEA training set. As before we generated a set of 20 reconstruction samples $\{\*x_i\}$ for our accelerated test images using all investigated techniques and obtained the corresponding segmentation $\{\*s_i\}$ using the segmentation net. We calculated the spread of the segmentation distribution using $\gamma$-maps which were defined by Baumgartner et al.~\cite{phiseg} as $\gamma(\{\*s_i\}) = \mathbb{E}[\text{CE}(\bar{\*s}, \*s_i)]$, where $\text{CE}$ denotes the cross-entropy and $\bar{\*s}$ is the mean segmentation. Fig.\,\ref{fig:var_segms} shows pairs of $\gamma$-maps and segmentation error maps for an example image in the OOD setting with 16x acceleration. The scaling of the color maps is shared for all images. Qualitatively samples generated by PHIRec exhibited an excellent correlation between the error and the variance outperforming the baselines. We also computed the NCC between the error maps and the segmentation variance (i.e. $\gamma$-maps) to measure the calibration of the segmentation uncertainty. The results are shown in column five of Fig.\,\ref{fig:metrics}. Again, PHiRec clearly outperformed the baseline methods. 

\begin{figure}[t]
    \centering
    \includegraphics[width=0.95\textwidth]{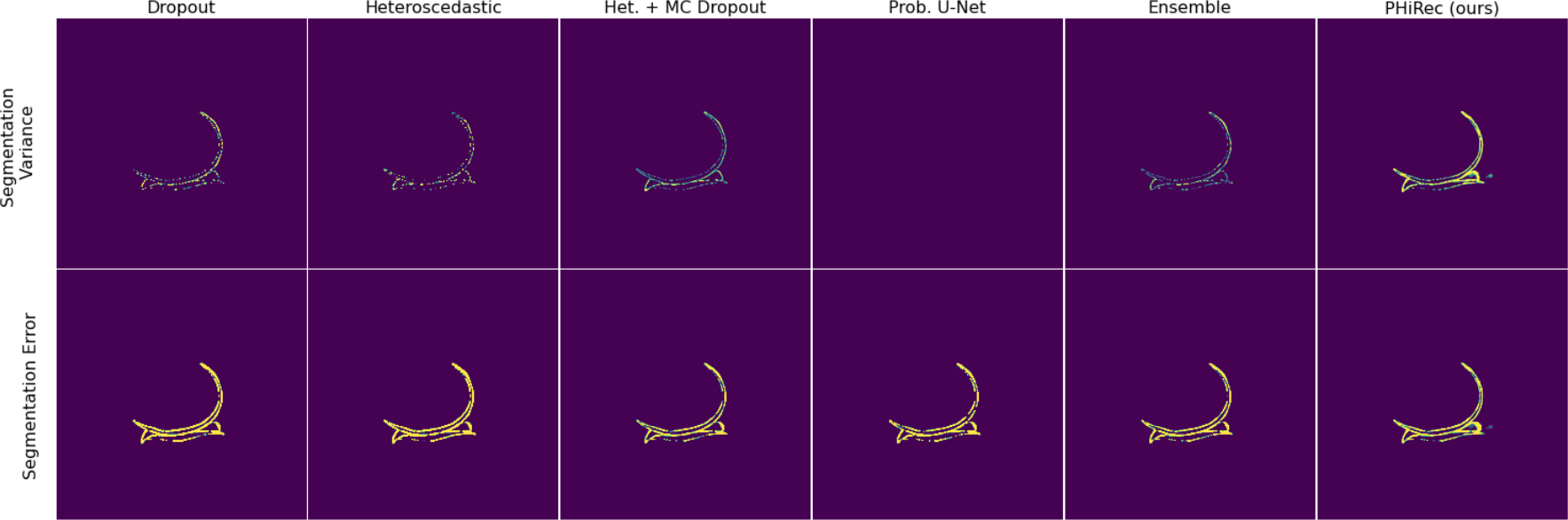}
    \caption{Segmentation variance maps (measured by $\gamma$-maps) and segmentation error maps in the OOD setting with 16x acceleration.}
    \label{fig:var_segms}
\end{figure}

\section{Discussion}

Well-calibrated uncertainty estimation is a crucial component for safely applying DL-based techniques to MRI reconstruction. In this paper, we described PHiRec, a novel reconstruction approach based on hierarchical conditional VAEs, which produces uncertainty estimates substantially better calibrated than several strong baselines. We further demonstrated, how uncertainties originating in the reconstruction process can be propagated to the downstream task of segmentation. In addition to our methodological contributions, we also present the, to our knowledge, first thorough quantitative comparison of different methods for uncertainty quantification in MRI reconstruction. 

Propagation of uncertainty to downstream tasks may allow to build fail-safe mechanisms to identify when uncertainties are too large to safely make a clinical decision or to guide a treatment. While our study used the simple U-Net as a base architecture and did not enforce data consistency with the measured $k$-space data, in future work we aim to combine our findings with state-of-the-art methods that are using multiple prediction and data consistency stages (e.g. \cite{schlemper2017deep,sriram_end--end_2020,kobler2020total}).  Future work will also focus on investigating the interplay of aleotoric and epistemic uncertainty for larger domain shifts such as changes in anatomy. 

\section*{Acknowledgments}
Funded by the Deutsche Forschungsgemeinschaft (DFG, German Research Foundation) under Germany’s Excellence Strategy – EXC number 2064/1 – Project number 390727645. The authors thank the International Max Planck Research School for Intelligent Systems (IMPRS-IS) for supporting Paul Fischer.

%
% ---- Bibliography ----
%
% BibTeX users should specify bibliography style 'splncs04'.
% References will then be sorted and formatted in the correct style.
%

\bibliographystyle{splncs04}
\bibliography{bibliography}
\newpage

\begin{center}
{\Large \bf Supplemental Materials \\ \vspace{2mm}
%Uncertainty Estimation and Propagation in \\ \vspace{1mm}
%Accelerated MRI Reconstruction}
}
\end{center}

\setcounter{figure}{0} 

\begin{figure}[!ht]
    \centering
    \includegraphics[width=\textwidth]{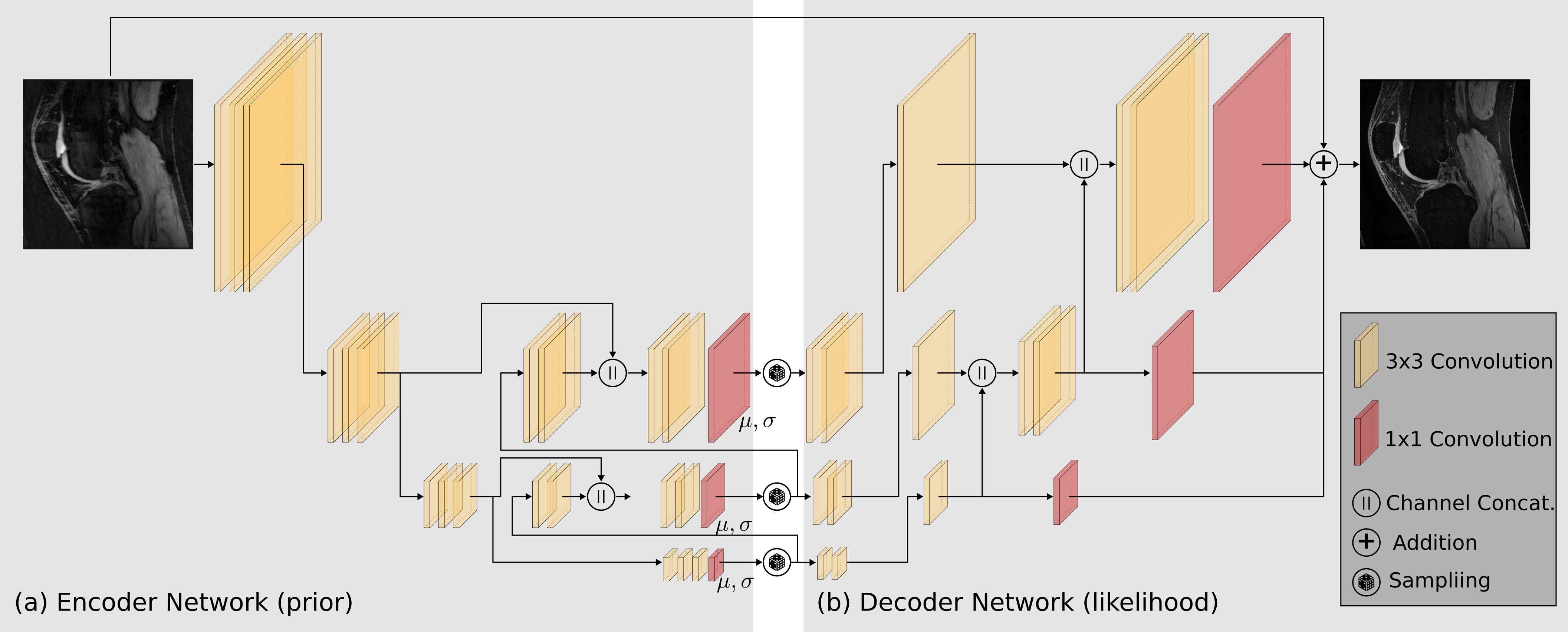}
    \caption{An overview of the proposed PHiRec network architecture as it is used for inference. Three latent levels $L=3$ are shown, but $L=5$ were used in all experiments. (a) The priors $p(\*z_l|\*x_u)$ are modelled using the encoder network. During training the posterior $q(\*z_l|\*x,\*x_u)$ is modelled using an identical encoder architecture but with an additional input channel for $\*x$. (b) The likelihood $p(\*x|\*z_{1:L}, \*x_u)$ is modelled using the decoder network. }
    \label{fig:phirec}
\end{figure}

\begin{figure}[h!]
    \centering
    \includegraphics[width=\textwidth]{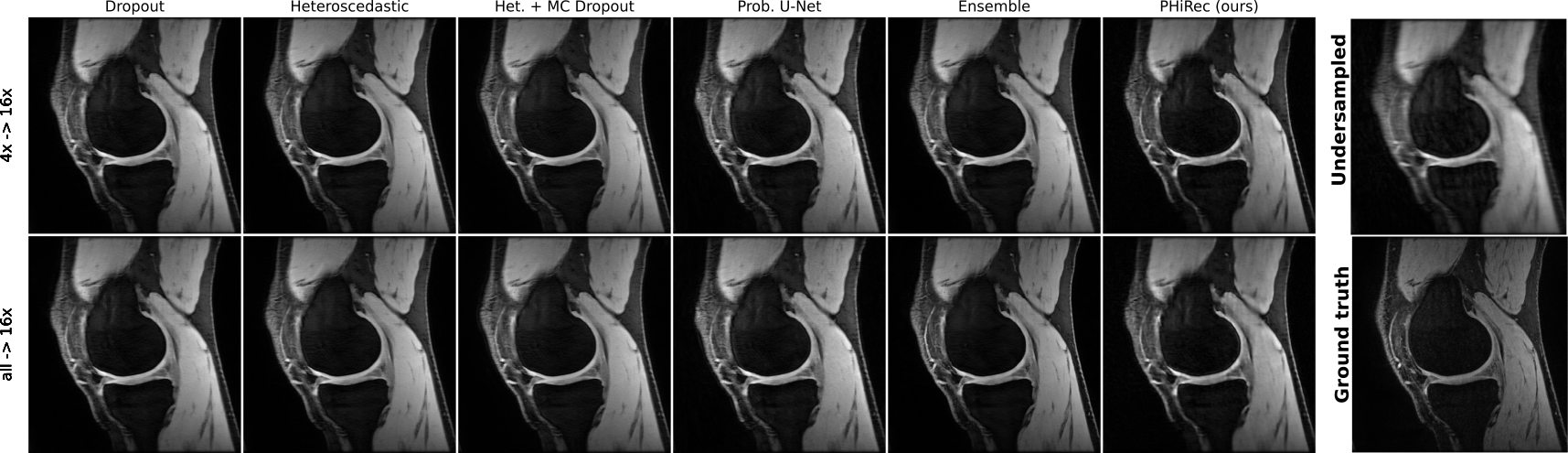}
    \caption{Example reconstructions for all methods in the in-domain (ID) and out-of-domain (OOD) setting at 16x acceleration. The right most column shows the undersampled input $\*x_u$ and the ground truth fully sampled image.}
    \label{fig:recon-collage}
\end{figure}

\begin{figure}[h!]
    \centering
    \includegraphics[width=\textwidth]{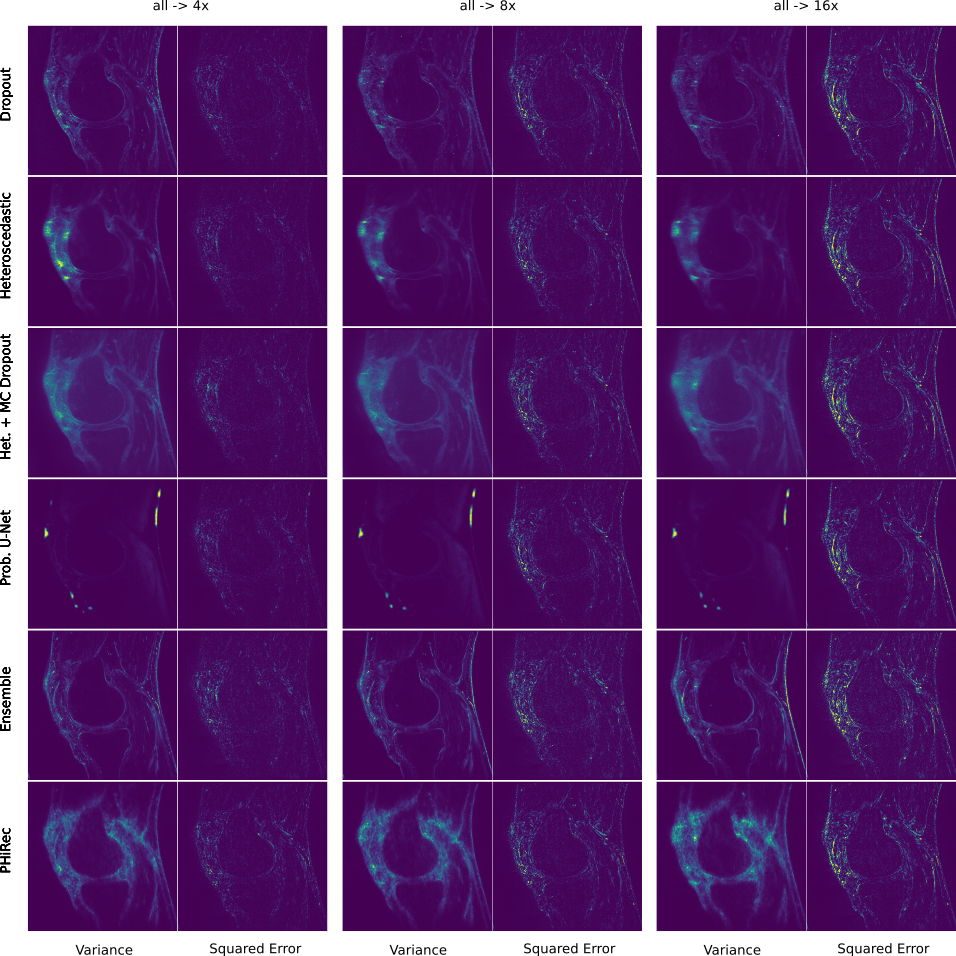}
    \caption{Variance maps and reconstruction squared error maps in the ID setting where the column labels have the format ``train acceleration'' \textrightarrow \ ``test acceleration''.}
    \label{fig:id-images}
\end{figure}

\end{document}